\documentclass[aps,prl,twocolumn,showpacs,superscriptaddress]{revtex4} 

\usepackage{amsfonts}
\usepackage{amsmath}
\usepackage{amssymb}
\usepackage{graphicx}
\usepackage{float}
\usepackage{bm}

\begin{document}

%%%%%%%%%%%%%%%%%%%%%%%%%%%% TITLE

\title{ Direct imaging of the structural domains in iron pnictides $A$Fe$_2$As$_2$ ($A$= Ca, Sr, Ba)}

%%%%%%%%%%%%%%%%%%%%%%%%%%%% AUTHORS

\author{M.~A.~Tanatar}
\affiliation{Ames Laboratory, Ames, Iowa 50011, USA}

\author{A.~Kreyssig}
\affiliation{Ames Laboratory, Ames, Iowa 50011, USA}
\affiliation{Department of Physics and Astronomy, Iowa State University, Ames, Iowa 50011, USA }

\author{S.~Nandi}
\affiliation{Ames Laboratory, Ames, Iowa 50011, USA}
\affiliation{Department of Physics and Astronomy, Iowa State University, Ames, Iowa 50011, USA }

\author{N.~Ni}

\affiliation{Ames Laboratory, Ames, Iowa 50011, USA}
\affiliation{Department of Physics and Astronomy, Iowa State University, Ames, Iowa 50011, USA }

\author{S.~L.~Bud'ko}
\affiliation{Ames Laboratory, Ames, Iowa 50011, USA}
\affiliation{Department of Physics and Astronomy, Iowa State University, Ames, Iowa 50011, USA }

\author{P.~C.~Canfield}
\affiliation{Ames Laboratory, Ames, Iowa 50011, USA}
\affiliation{Department of Physics and Astronomy, Iowa State University, Ames, Iowa 50011, USA }

\author{A.~I.~Goldman }
\affiliation{Ames Laboratory, Ames, Iowa 50011, USA}
\affiliation{Department of Physics and Astronomy, Iowa State University, Ames, Iowa 50011, USA }

\author{R.~Prozorov}
\email[Corresponding author: ]{prozorov@ameslab.gov}
\affiliation{Ames Laboratory, Ames, Iowa 50011, USA}
\affiliation{Department of Physics and Astronomy, Iowa State University, Ames, Iowa 50011, USA }

\date{\today}

%%%%%%%%%%%%%%%%%%%%%%%%%%%% ABSTRACT

\begin{abstract}
The parent compounds of recently discovered iron-arsenide superconductors, $A$Fe$_2$As$_2$ with alkaline earth $A$=Ca, Sr, Ba, undergo simultaneous structural and magnetic phase transitions at a temperature $T_{SM}$. Using a combination of polarized light microscopy and spatially resolved high-energy synchrotron X-ray diffraction we show that the orthorhombic distortion leads to the formation of 45$^o$-type structural domains in all parent compounds. Domains penetrate through the sample thickness in the $c$- direction and are not affected by crystal defects such as growth terraces. The domains form regular stripe patterns in the plane with a characteristic dimension of 10 to 50 $\mu m$. The direction of the stripes is fixed with respect to the tetragonal (100) and (010) directions but can change by 90$^o$ on thermal cycling through the transition. This domain pattern may have profound implications for intrinsic disorder and anisotropy of iron arsenides.
\end{abstract}


%Metals, transport processes in, 72.15.-v

%Superconducting materials 74.70.Dd Ternary, quaternary, and multinary compounds (including Chevrel phases, borocarbides, etc.)
%68.37.-d Microscopy of surfaces, interfaces and thin films
%61.50.Ks Crystallographic aspects of phase transformations, pressure effects

% 61.05.cp X-ray diffraction

\maketitle

%%%%%%%%%%%%%%%%%%%%%%%%%%%% INTRODUCTION

%%%%1) Introduction
%Oxypnictide superconductors, discovery, properties

%\section{Introduction}

Until last spring, there was only one class of high transition temperature (high-$T_c$) superconductors - the cuprates \cite{cuprates}. Discovery of superconductivity \cite{Hosono} with $T_c$ currently as high as 56 K \cite{Nd,Sm} in iron arsenide compounds $R$FeAs(O,F) (abbreviated as $R$-1111 with $R$ = rare earth), has fueled hopes that a new chemical and physical perspective can shed light on the nature of the high-temperature superconductivity. 

Comparison of the cuprates and iron arsenides indeed reveals some similarities. Both classes of compounds have well defined layered structures. Within the layers the transition metals Cu and Fe are arranged on a square lattice and their orbitals make the main contribution to the electronic density of states at the Fermi level \cite{MP1}. However, the electronic structures of the parent compounds are notably different: the cuprates are magnetically ordered Mott insulators, whereas in the iron arsenides the magnetic transition occurs in the metallic state below $T_M$ and is accompanied by a structural distortion below $T_S$. In both cases the superconductivity appears on doping of the parent compounds, suppressing the magnetic order. The character of the magnetic order is notably different as well. In the cuprates, the magnetism is characterized by a simple antiferromagnetic arrangement of spins in the tetragonal planes \cite{Vakhnin} and is very anisotropic (close to two-dimensional) \cite{Birgeneau}. In the iron arsenides, the spins on the iron sites form stripe order in the basal planes, with ferromagnetic spin alignment within the stripes and antiferrmognetic order between stripes both in the plane and perpendicular to the plane. \cite{1111strmagn}

The stripe type of magnetic order is incompatible with the tetragonal symmetry of the crystal. Its formation becomes possible because the crystal undergoes a tetragonal-to-orthorhombic structural transition. This transition precedes magnetic ordering in the parent $R$-1111 compounds \cite{1111strmagn} ($T_S > T_M$), whereas both transitions occur simultaneously in $A$Fe$_2$As$_2$ with $T_S$=$T_M$=$T_{SM}$. \cite{sim_order_122Ca,sim_order_122Sr,sim_order_122Ba} 
It is not yet clear whether magnetism or structural distortion is the driving force behind the transition.

In this paper we show that the orthorhombic structural phase transition leads to the formation of a pattern of structural twin domains. These domains are thin  plates with the extended surface parallel to the tetragonal $c-$axis and they span the entire crystal. We determined that the domains are of the 45 degree type, with faces corresponding to the (110) and (1$\bar{1}$0) planes in the orthorhombic notation, as expected for the distortion along the orthorhombic $a$-axis. This is similar to the structural domains formed during growth in the orthorhombic phase of YBa$_2$Cu$_3$O$_7$.\cite{ZPhys1988,domains_neutrons} 

%\section{Experimental}

Single crystals of BaFe$_2$As$_2$ and of SrFe$_2$As$_2$ were grown from FeAs flux from a starting load of metallic Ba (or Sr) and FeAs, as described in detail elsewhere. \cite{NiNiCo,SrFe2As2_growth} Crystals were thick platelets with sizes as big as 12$\times$8$\times$1 mm$^3$ and large faces corresponding to the tetragonal (001) plane. For polarized optical imaging, the samples were cleaved with a razor blade into rectangular shaped platelets of typically (2-5)$\times$(2-5) $\times$ (0.1-0.5) mm$^3$. Single crystals of CaFe$_2$As$_2$ were grown from Sn flux as described elsewhere. \cite{sim_order_122Ca}  The crystals were the same size as crystals of Ba and Sr compounds, but due to both very clean natural growth faces corresponding to (001) and (011) crystallographic planes and difficulty to form a good cleave surface they were used as-grown in the study. 

X-ray diffraction measurements were performed to characterize the quality of the single crystals. The absorption length of the high energy (99.6~keV) x-rays from the synchrotron source (beamline 6ID-D in the MUCAT sector at the Advanced Photon Source, Argonne) was about 1.5~mm, which allowed the study of the entire sample volume. With the use of a slit system, the 

\begin{figure}[H]
\begin{center}
\includegraphics[width=8.5cm]{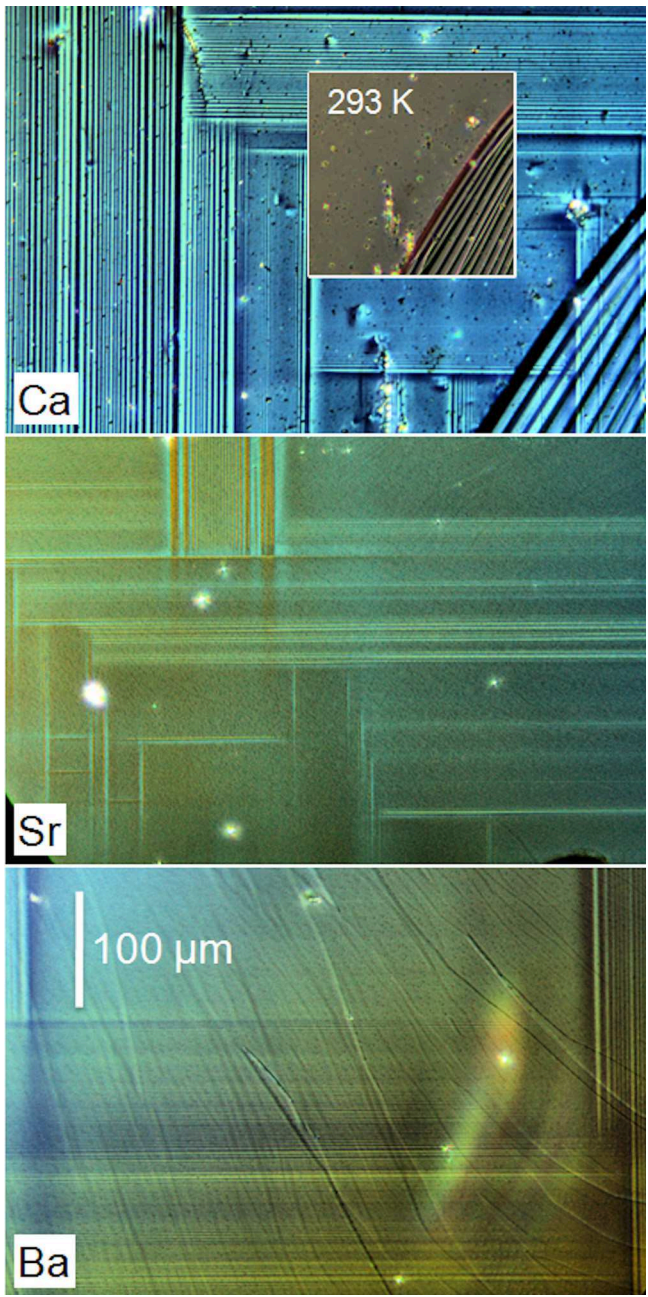}
 \end{center}
\caption{(Color online) A pattern of structural domains is found in parent compounds of iron-pnictide superconductors, $A$Fe$_2$As$_2$, with $A$=Ca (top panel), Sr (middle panel) and Ba (bottom panel) at 5~K, well below $T_{SM}$, the temperature of the simultaneous structural/magnetic transition. The characteristic spacing between the lines is about 10 $\mu$m, and the contrast in optical images follows the magnitude of orthorhombic distortion in the compound. Inset in top panel shows terraces on $A$=Ca crystal at room temperature. In all images $c$-axis is perpendicular to the surface. }
	\label{domains_lowT}
\end{figure}

\noindent 
incident beam was reduced to a 0.1x0.1~mm$^2$ area. The samples were mounted with their {\bf c}-direction parallel to the incident beam and the images were scanned performing stepwise translations of the sample perpendicular to the incident beam. The entire sample was scanned in both directions and no impurities, misoriented grains or diffuse signals from disordered material were detected.
Therefore, from the x-ray study we conclude that the entire sample is a high-quality single crystal. Two-dimensional scattering patterns were measured by a MAR345 
image-plate positioned 1705 mm behind the sample. The direct beam was blocked by a beam stop in front of the detector. The sample was rocked by $\pm$2.4$^o$ about both axes perpendicular to the incoming beam to record complete planes of Bragg reflections as described in Ref.~\onlinecite{KreyssigPRB07}.

White - light optical images were taken in a polarization microscope {\it Leica DMLM} with polarizer and analyzer in almost crossed position. A helium flow cryostat was positioned on the x-y stage of the microscope and allowed direct imaging from room temperature down to 5 K. High resolution static images as well as real time video were recorded. The contrast of the domain images is higher for larger difference in the rotation of the polarization plane between neighboring domains, proportional to the degree of orthorhombic distortion, $(a_O - b_O)/ (a_O + b_O)$.

In Fig.~\ref{domains_lowT} we show optical images of single crystals of the 122 compounds, taken at 5~K, well below $T_{SM}$ of 173~K (Ca) \cite{sim_order_122Ca}, 205~K (Sr) \cite{sim_order_122Sr} and 137~K (Ba) \cite{sim_order_122Ba}.  A regular pattern of domain boundaries oriented in two orthogonal directions is clearly visible in all cases. The contrast of the pattern is highest in the $A$=Ca compound, consistent with the largest orthorhombic distortion, equal to 0.67\% for $A$=Ca,\cite{sim_order_122Ca} 0.55 \% for $A$=Sr \cite{sim_order_122Sr} and  0.36 \% for $A$= Ba. \cite{sim_order_122Ba} A typical domain width is about 10~$\mu$m. Over large areas, sometimes covering the whole surface of the crystal, domains form stacks of parallel plates. In some areas perpendicular domain sets inter-penetrate. A supplementary material real-time movie shows the evolution of the domain walls in a warming-cooling cycle \cite{movie}. The process of domain formation shows pronounced hysteresis with temperature, and the direction of the dominant lamellae pattern can change by 90 degrees from one run through $T_{SM}$ to another. This is notably different to the case of YBCO, where the domains are \textquotedblleft  pinned \textquotedblright by the defects even after a process of deoxigenation-reoxigenation \cite{YBCO-domain_pinned}.

The lamellae are not affected by the macroscopic crystal defects. As can be seen in the bottom-right corner in the top panel in the Fig.~\ref{domains_lowT} for the Ca compound, the crystal under study has terraces on the sample surface (running at an angle to the figure frame and shown in inset at room temperature), with a step size of the order of 20 $\mu$m along the $c$-axis. On crossing the terraces, the lamellae lines perfectly match at different levels. This clearly shows that the domain walls are extended along the $c$-axis, as expected for a tetragonal-to-orthorhombic distortion. We should emphasize that the domain boundaries are very smooth and regular, revealing the very high  

\begin{figure} [H]
	\includegraphics[width=8.5cm]{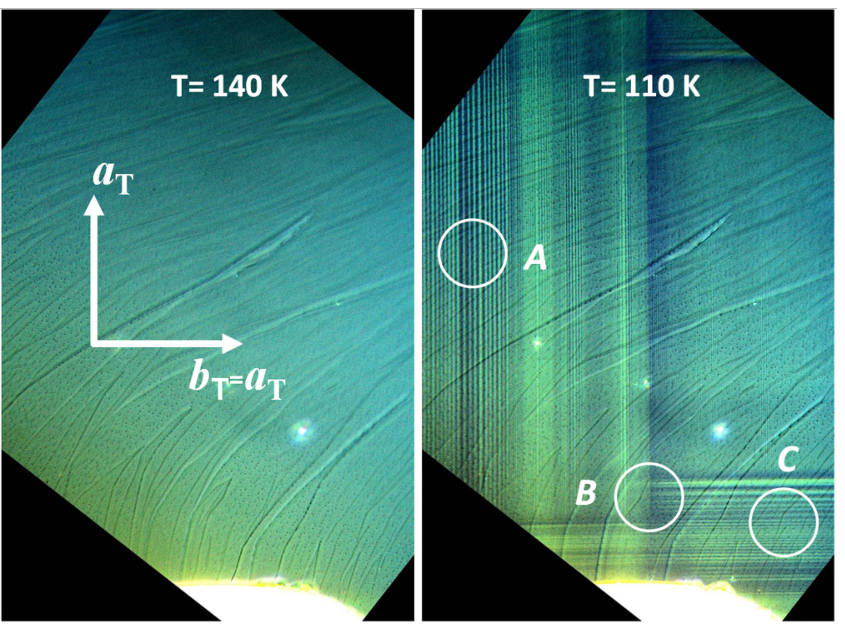}
	\includegraphics[width=8.5cm]{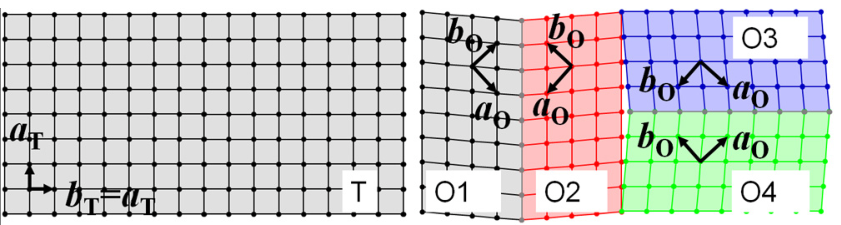}
	\includegraphics[width=8.5cm]{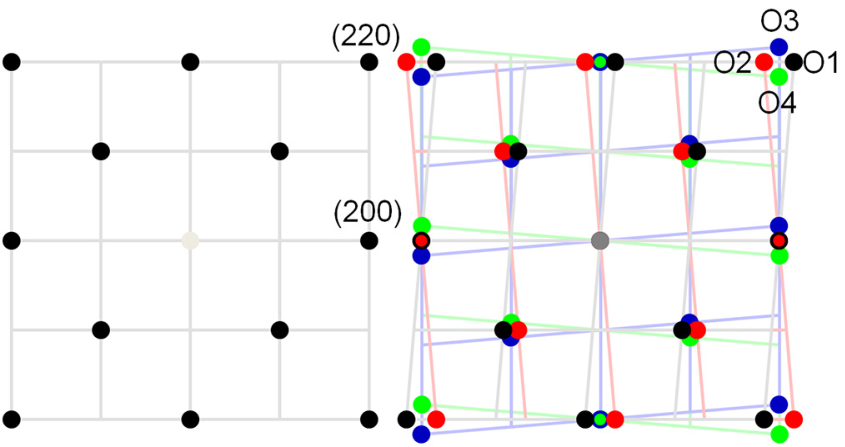}
	\includegraphics[width=8.5cm]{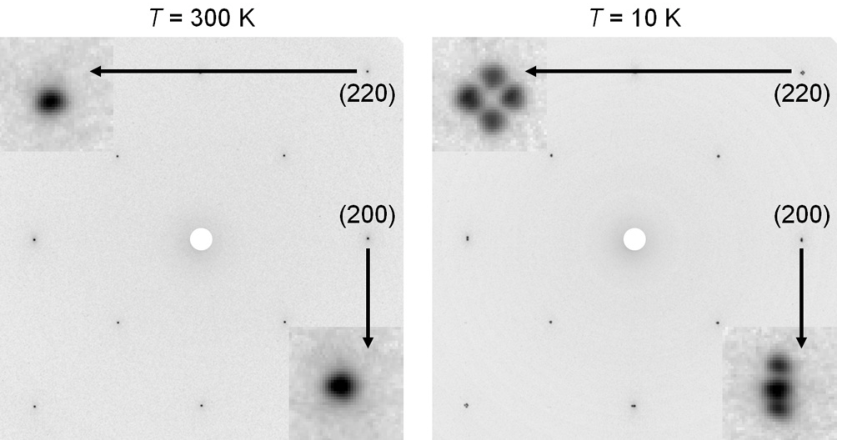}
\caption{(Color online) High resolution optical image of pure BaFe$_2$As$_2$ above (left) and below (right) the temperature of the coupled structural/magnetic transition, $T_{SM}$=135~K. A pattern of domain walls is formed below $T_{SM}$ due to formation of twin boundaries (top row). Second row of panels shows schematics of the displacements of atoms in the tetragonal lattice (above $T_{SM}$) during structural transition, leading to orthorhombic distortion and formation of domain walls. In the orthorhombic phase the unit cell of the lattice doubles in size in the $a-b$ plane, new unitary vectors $a_O$ and $b_O$ become of different length ($a_O > b_O$), and rotate by approximately 45 degrees. The third row shows sketches of the expected transformation of the X-ray diffraction pattern during domain formation. The bottom raw shows actual X-ray data and zooms of (200) and (220) reflections (insets). }
	\label{mechanism}
\end{figure}

\noindent
quality of the crystals used. On the cleaved surfaces of the Sr and Ba compounds, the domains have smaller spacing than in the Ca compound.

\begin{figure} [tb]
	\includegraphics[width=8.5cm]{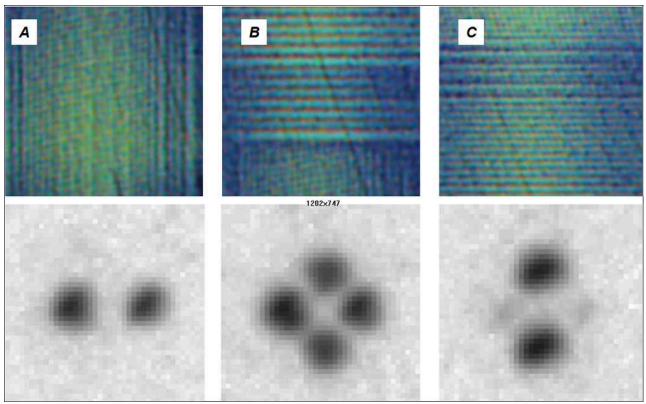}
	\caption{(Color online) Below the transition, the domain boundaries form regular patterns. Three typical areas can be found in both optical images and x-ray microdiffractograms around the split tetragonal (220) reflection. The two rows compare typical patterns as found in optical and x-ray images. Of note, since x-ray penetrate the whole sample thickness, the lack of crossed patterns in left and right panels shows clearly that domains are extending through the whole thickness of the samples along (001) axis. The area $A$ represents lamellae of the domains O1 and O2, the area $C$ represents lamellae of domains O3 and O4. The area $B$ shows crossed lamellae of the domain walls O1-O2 and O3-O4. It is characterized by increased structural deformation.}
	\label{areas}
\end{figure}

The top panel of Fig.~\ref{mechanism} shows optical images of the BaFe$_2$As$_2$ single crystal above (140~K) and below (110~K) $T_{SM}$=137~K. Several defects at the surface and sample edges can be used as markers. The domains occupy most of the crystal surface, and run along two perpendicular

\noindent
 directions, see Fig.~\ref{areas}, spots A and C. In some areas the two domain patterns intersect each other, spot B.

In the second row of panels in Fig.~\ref{mechanism} we show the schematics of the domain wall formation during orthorhombic lattice distortion. In the tetragonal lattice, characterized by the unit cell vectors $a_T$ and $b_T = a_T$ within the plane, the atoms occupy positions at the nodes (left panel). Orthorhombic distortion leads to atomic displacement along the diagonal of the tetragonal lattice, with $a_O>b_O$, so that the orthorhombic unit cell rotates by approximately 45$^o$ and doubles in size (right panel). The displacements of atoms along the unit vectors of the orthorhombic unit cell can lead to four different domain patterns (O1 to O4) (different colors online). Similar patterns of domains were studied extensively in the orthorhombic phase of YBCO. \cite{ZPhys1988} Since domains O1 and O2 [O3 and O4] share a common plane corresponding to the tetragonal (100) [(010)] plane, their formation does not require lattice deformation and they easily form pairs. A boundary between pairs O1-O2 and O3-O4 is heavily distorted and the areas of mismatch are usually characterized by atomic displacements from regular positions \cite{ZPhys1988} and strong stress and strain.

The cause of domain formation is the simultaneous nucleation of the low temperature phase below $T_{SM}$ in a number of points in the sample. The orthorhombic distortion makes it energetically favorable to form domains, since they release stress over the entire lattice in small areas of high deformation. As can be seen from Fig.~\ref{mechanism}, the directions of the orthorhombic distortion and of the domain boundaries form an angle of about 45$^o$, with domain wall corresponding to (110) [or (1$\bar{1}$0)] crystallographic planes. 

The x-ray diffraction pattern for temperatures above and below the transition reveals a splitting of the (220) spot into four spots, while each (200) spot splits into three, with the direction of splitting corresponding to the direction of twin boundary in the reciprocal space. The two bottom rows in Fig.~\ref{mechanism} show the effect of domain formation on the x-ray patterns in reciprocal space and the actual x-ray diffraction patterns for BaFe$_2$As$_2$ focusing on the (220) and (200) spots. 
This correspondence is illustrated in further detail in Fig.~\ref{areas}. Three panels show x-ray diffraction pattern obtained with 100*100 $\mu m^2$ spatial resolution and corresponding orientation of structural domains with respect to the $ab$ plane in real space. Of note, since high-energy x-ray diffraction probes the whole sample thickness, observation of singular pairs of (220) spots shows that the domains extend along $c$-crystallographic direction through the whole sample thickness. This shows that domains are true bulk phenomenon.

%%%%%%%%%%%%%%Fig two-panel, x-ray domains and domain meshes 

%\section{Discussion}

A recent transmission electron microscopy (TEM) study of domain boundaries in 122 parent compounds found domains which were notably smaller in size than in our study, typically 100 to 400 nm \cite{TEM}. Even smaller, about 40 nm, domains were reported for the Ca compound above $T_{SM}$. In our study we can not resolve the domains of this size, however, in big sample areas where domains themselves were not spatially resolved, passing through $T_{SM}$ clearly changed the brightness of the image. We do not see any change in the contrast of high quality single-domain areas of Ca-122 samples grown from Sn flux on warming above $T_{SM}$. Since the crystals used in the TEM study were grown from self-flux, which for the Ca compound are of lower quality \cite{Ca_FeAsflux}, this difference can be caused by sample preparation technique. Indirectly, this is supported by the comparison of the sharpness of the resistivity jump at $T_{SM}$, measured in our sample \cite{anisotropy_pure} and Ref.~\onlinecite{TEM}. On the other hand sample milling required for TEM can lead to notable stress in the sample, which can be another cause of the difference. 

The anisotropy of the electronic structure is believed to be an important parameter for understanding superconductivity in iron arsenides \cite{Mazin,Chubukov}. Measurements of the anisotropy of electrical resistivity in parent compounds found systematically small values of anisotropy \cite{anisotropy_pure}, both above and below $T_{SM}$ and the slight anisotropy decreases on passing the transition. Since domains penetrate the whole sample along the $c$-axis, it is reasonable to assume that they will affect less the current flow along the $c$-axis, and thus $\rho_c$ measurements. Measurements of in-plane resistivity, $\rho_a$, on the other hand, should represent an average of $\rho_a$ and $\rho_b$, similar to twinned crystals of YBCO \cite{Tozer}. In addition, the value of in-plane resistivity can be notably increased by boundary scattering. Since rapid decrease of resistivity is observed in the Sr and Ba 122 parent compounds below $T_{SM}$, this suggests that domains are rather transparent for electrons.

%\section{Conclusions}

In conclusion, the parent compounds of iron arsenic superconductors reveal clear pattern of twin domains below the temperature of tetragonal to orthorhombic transition. The domain walls run at approximately 45$^o$ to the orthorhombic lattice and span the whole sample along $c$-axis direction.

We thank Doug Robinson for the support of the high-energy x-ray measurements.
Work at the Ames Laboratory and at the MUCAT sector was supported by the Department of Energy-Basic Energy Sciences under Contract No. DE-AC02-07CH11358. The use of the Advanced Photon Source was supported by the U.S. DOE under Contract No. DEAC02-06CH11357. M.A.T. acknowledges continuing cross-appointment with the Institute of Surface Chemistry, National Ukrainian Academy of Sciences. R. P. acknowledges support from Alfred P. Sloan Foundation.

%%%%%%%%%%%%%%%%%%%%%%%%%%%% BIBLIOGRAPHY


\begin{references}
\bibitem{cuprates} J. G. Bednorz and K. A. Muller, Rev. Mod. Phys. {\bf 60}, 585 (1988).

\bibitem{Hosono} Y. Kamihara, T. Watanabe, M. Hirano, and H. Hosono, J. Am. Chem. Soc. {\bf 130}, 3296 (2008).

\bibitem{Nd} Z.~A.~Ren, J.~Yang, W.~Lu, W.~Yi, X.~L.~Shen, Zh.-C.~Li, G.-C.~Che, X.~L.~Dong, L.~L.~Sun, F.~Zhou, Z.-X. Zhao, Eur. Phys. Lett. {\bf82}, 57002 (2008).

\bibitem{Sm} X.~H.~Chen, T.~Wu, G.~Wu, R.~H.~Liu, H.~Chen, D.~F.~Fang, Nature {\bf  453}, 761 (2008).

\bibitem{MP1} I. Mazin, M. D. Johannes, L. Boeri, K. Koepernik, and D. J. Singh, Phys. Rev. B {\bf 78}, 085104 (2008).

\bibitem{Vakhnin} D. Vaknin, S. K. Sinha, D. E. Moncton, D. C. Johnston, J. M. Newsam, C. R. Safinya, and H. E. King, Jr., Phys. Rev. Lett. {\bf 58}, 2802 (1987).

\bibitem{Birgeneau} 
G. Shirane, Y. Endoh, R. J. Birgeneau, M. A. Kastner,
Y. Hidaka, M. Oda, M. Suzuki, and T. Murakami
Phys. Rev. Lett. {\bf 59}, 1613 (1987).

\bibitem{1111strmagn}
C. de la Cruz, Q. Huang, J.~W.~Lynn, J.~Y.~Li, W. Ratcliff, J. ~L.~Zarestky, H.~A.~Mook, G.~F.~Chen, J.~L.~Luo, N.~L.~Wang, P.~C.~Dai, Nature {\bf  453}, 899 (2008).

\bibitem{sim_order_122Ca}
N. Ni, S.~Nandi, A.~Kreyssig, A.~I.~Goldman, E.~D.~Mun, S.~L.Bud'ko, and P.~C.~Canfield, Phys. Rev. B {\bf 78}, 014523 (2008); A. I. Goldman, D. N. Argyriou, B. Ouladdiaf, T. Chatterji, A. Kreyssig, S. Nandi, N. Ni, S.~L.Bud'ko, 
P. C. Canfield, and R. J. McQueeney, {\it ibid} {bf 78} 100506 (2008).

\bibitem{sim_order_122Ba}
M.~Rotter, M.~Tegel, D.~Johrendt, I.~Schellenberg, W.~Hermes, and R.~Pottgen
Phys Rev. B {\bf 78}, 020503 (2008).

\bibitem{sim_order_122Sr}
M. Tegel, M. Rotter, V. Weiss, F.~M.~ Schappacher , R.~ Poettgen, and D.~Johrendt, J. Phys. Cond. Mat. {\bf 20}, 452201 (2008).


\bibitem{ZPhys1988} H.~Schmid, E.~Burkhardt, E.~Walker, X.~Brixel, M.~Clin, J.~P.~Rivera, J.~L.~Jorda, M. Francois, and K.~Yvon, Z. Phys. B {\bf 72}, 305 (1988).

\bibitem{domains_neutrons} G. J. McIntyre, A. Renault, and G. Collin, Phys. Rev. B {\bf 37},
5148 (1988).

\bibitem{NiNiCo} N. Ni, M. E. Tillman, J. Q. Yan, A. Kracher, S. T. Hannahs, S. L. Bud'ko, and P. C. Canfield, Phys. Rev. B {\bf 78} 214515 (2008) 

\bibitem{SrFe2As2_growth}
J.~Q.~Yan, A.~Kreyssig, S. ~Nandi, N.~Ni, S.~L.~Bud'ko, A.~Kracher, R.~J. McQueeney, R.~W.~McCallum, T.~A.~Lograsso, A.~I.~Goldman, and P.~C. Canfield
Phys. Rev. B {\bf 78}, 024516 (2008).


\bibitem{KreyssigPRB07} A.~Kreyssig, S. Chang, Y. Janssen, J.~W. Kim, S. Nandi, J. Q. Yan, L. Tan, R. J. McQueeney, Phys. Rev. B {\bf 76}, 054421 (2007). 

\bibitem{YBCO-domain_pinned}
V.~I.~Voronkova, and T. Wolf, Physica C {\bf 218}, 175 (1993).

\bibitem{TEM} C. Ma, H. X. Yang, H. F. Tian, H. L Shi, J. B. Lu, Z. W. Wang, L. J. Zeng, G. F. Chen, N. L. Wang, and J. Q. Li, Phys. Rev. B {\bf 78}, 060506 (R) (2009). 

\bibitem{Ca_FeAsflux} A. I. Goldman, A. Kreyssig, K. Prokes, D. K. Pratt, D. N. Argyriou, J. W. Lynn, S. Nandi, S. A. J. Kimber, Y. Chen,  Y. B. Lee, G. Samolyuk, J. B. Le\~{a}o, S. J. Poulton, S. L. Bud'ko, N. Ni, P. C. Canfield, B. N. Harmon,  and R. J. McQueeney Phys. Rev. B {\bf 79}, 024513 (2009). 

\bibitem{anisotropy_pure} M. A. Tanatar, N. Ni, G. D. Samolyuk, S. L. Bud'ko, P. C. Canfield, and R. Prozorov, Arxiv 0903.0820

\bibitem{Mazin} I.~I.~Mazin, D.~J.~Singh, M.~D.~Johannes, M.~H.~Du, 
Phys. Rev. Lett. {\bf 101}, 057003 (2008).

\bibitem{Chubukov} A.~V.~Chubukov, D.~V.~Efremov, and I.~Eremin, Phys. Rev. B {\bf 78}, 134512 (2008).

\bibitem{Tozer}S. W. Tozer, A. W. Kleinsasser, T. Penney, D. Kaiser, and F.
Holtzberg, Phys. Rev. Lett. {\bf 59}, 1768 (1987).

\bibitem{movie} Online supplementary video. Also, see http://www.cmpgroup.ameslab.gov/supermaglab/video /Pnictides.html


%\bibitem{anisotropy} M. A. Tanatar, N. Ni, C. Martin, R. T. Gordon, H. Kim, V. G. Kogan, G. D. Samolyuk, S. L. Bud'ko, P. C. Canfield, and R. Prozorov, Phys. Rev. B in press.

%\bibitem{ProzorovPhysicaC} R. Prozorov, M. A. Tanatar, R. T. Gordon, C. Martin, H. Kim, V. G. Kogan, N. Ni, M. E. Tillman, S. L. Bud'ko, and P. C. Canfield, arXiv0901.3698 (unpublished).


\end{references}
\end{document}